\documentclass[twocolumn,aps,prc,showpacs,superscriptaddress,floatfix]{revtex4-2}
\usepackage{rotating}
\usepackage{epsfig} 
\usepackage{hyperref}
\usepackage{color}
\usepackage{url}
\usepackage{multirow}

\usepackage{amsmath}
\usepackage{lineno}
\usepackage{graphicx}
\usepackage[caption=false]{subfig}

\graphicspath{{./}, {./image/}}
\DeclareGraphicsExtensions{{.pdf}}

\makeatletter
\renewcommand{\p@subsection}{}
\renewcommand{\p@subsubsection}{}
\makeatother

\begin{document}

\newcommand{\snn}{\sqrt{s_{\textsc{nn}}}}
\newcommand{\hijing}{{\textsc{hijing}}}
\newcommand{\AuAu}{Au+Au}
\newcommand{\UU}{U+U}
\newcommand{\U}{{\rm U}}
\newcommand{\Nch}{N_{\rm ch}}
\newcommand{\pt}{p_T}
\newcommand{\Pt}{[\pt]}
\newcommand{\vpt}{$v_2$-$\pt$}

\newcommand {\mean}[1]   {\langle{#1}\rangle}
\newcommand {\mV} {\mean{V_2}}
\newcommand {\mv} {\mean{v_2^2}}
\newcommand {\mdpt} {\mean{(\delta\pt)^2}}
\newcommand {\mvpt} {\mean{v_2^2\delta\pt}}


\title{The nonflow issue in connecting anisotropy measurements to hydrodynamics in relativistic heavy-ion collisions}

\author{Fuqiang Wang}
\email{fqwang@purdue.edu}
\address{Department of Physics and Astronomy, Purdue University, West Lafayette, IN 47907}


\begin{abstract}
    Hydrodynamics can describe majority of the measured azimuthal anisotropies in relativistic heavy-ion collisions. Many of the anisotropy measurements are contaminated by nonflow correlations (i.e., those unrelated to global event-wise correlations). Those nonflow contamination can cause incorrectness or compromise the accuracy of the physics extracted from data-hydrodynamics comparison, particularly when one relies on subtle difference in the measurements. In the recent preprint by STAR (arXiv:2401.06625) extracting the Uranium nucleus deformation parameter, nonflow contamination is assessed by subevents in the limited STAR acceptance. In this note, we demonstrate that such assessment is inadequate and illustrate how large an effect nonflow can cause by using the \hijing\ model, in which all correlations are nonflow and non-hydrodynamic. We thereby conclude that the extracted Uranium deformation parameter is premature and emphasize the importance of an earnest assessment of or correction for nonflow contamination, not only for this STAR analysis but more generally for studies relying on comparing anisotropy measurements to hydrodynamic calculations.
\end{abstract}


\maketitle

\section{Introduction} \label{sec:introduction}
Anisotropic flow is a hallmark of heavy-ion (nucleus-nucleus) collisions. It arises from interactions converting the anisotropy in configuration space of a nucleus-nucleus collision into momentum space~\cite{Ollitrault:1992bk}. The configuration space anisotropy is prominent in non-central collisions where the overlap interaction zone is of an almond shape. The anisotropy is nonzero even in central head-on collisions because of position fluctuations of nucleons in the colliding nuclei, giving rise to finite eccentricities~\cite{Alver:2008zza}. Because of the same reason, nonzero anisotropy can also emerge in small-system collisions, such as proton-proton and proton-nucleus collisions~\cite{Li:2012hc,Nagle:2018nvi}.

The interactions in the system created in relativistic heavy-ion collisions, presumably the quark-gluon plasma (QGP), are governed by quantum chromodynamics (QCD). How exactly those interactions convert the initial-state geometry anisotropy into final-state momentum anisotropy is not well settled. It is generally believed that ultra-strong interactions are required to produce the observed large anisotropy (or flow) in midcentral to central heavy-ion collisions, and the QGP created in those collisions is a nearly perfect fluid~\cite{Gyulassy:2004zy} and can be well described by viscous hydrodynamics~\cite{Heinz:2013th}. In peripheral heavy-ion collisions and small-system collisions, the interactions may not be strong enough where hydrodynamics would be applicable and the escape mechanism may be at work~\cite{He:2015hfa,Romatschke:2015dha,Kurkela:2018qeb,Kurkela:2021ctp}.

Interactions also generate transverse momentum, absent in the initial state. Event-by-event correlations between transverse momentum and momentum anisotropy naturally emerge. For instance, in head-on collisions of spherical nuclei, a fluctuated smaller interaction zone result in stronger interactions causing positive correlations between elliptic anisotropy ($v_2$) and event-wise mean transverse momentum ($\Pt$). An interesting case is deformed nuclei like the Uranium (U): in central \UU\ collisions, tip-tip configuration yields a smaller and more spherical collision zone, hence larger $\Pt$ and weaker $v_2$, whereas body-body configuration gives a larger and less spherical collision zone, hence smaller $\Pt$ and stronger $v_2$; a \vpt\ anti-correlation results~\cite{Giacalone:2019pca,Giacalone:2021udy}. 

Because anisotropic flow is a result of interactions from an anisotropic initial geometry, studies of anisotropic flow can reveal important physics about the interactions and the initial geometry. Two types of studies can be taken: (1) with colliding nuclei of well measured  density distributions, one may gain information about the created QGP, such as its shear viscosity to entropy density ratio, via its hydrodynamic responses~\cite{Song:2010mg}; and (2) with well-controlled hydrodynamic responses, one may probe structures of the colliding nuclei and thereby the physics governing those structures~\cite{Li:2019kkh,Giacalone:2021udy}. 
Because the conversion from initial-state geometry anisotropy to final-state momentum anisotropy is dependent of the system evolution, significant uncertainties are inevitable~\cite{Teaney:2003kp,Niemi:2011ix}.

For studies of type (2), an effective strategy is to utilize a pair of nuclei close in mass number with yet appreciable difference in their density distributions (nuclear structure). One example is the isobar $^{96}_{44}$Ru+$^{96}_{44}$Ru and $^{96}_{40}$Zr+$^{96}_{40}$Zr collisions, conducted in 2018 at the Relativistic Heavy-Ion Collider (RHIC) to search for the chiral magnetic effect~\cite{Voloshin:2010ut,Kharzeev:2015zncReviewCME,Skokov:2016yrj,Zhao:2019hta,STAR:2021mii}. Density functional theory calculations revealed differences in the nuclear structures and neutron skins of the Ru and Zr nuclei, predicting significant differences in the final-state multiplicities and anisotropic flows between the isobar collisions~\cite{Xu:2017zcn,Li:2018oec,Li:2019kkh}, later confirmed by experiment~\cite{STAR:2021mii}. 

However, there is an important complication with anisotropic flow measurements (and similarly other correlation measurements). Flow anisotropies are often measured by two-particle correlations via 
\begin{equation}
    \mean{V_n} = \mean{\cos n(\phi_1-\phi_2)}\,,
    \label{eq:V2}
\end{equation}
where $\phi_1$ and $\phi_2$ are the azimuthal angles of two particles of interest (POI), and $n$ is the harmonic order (for instances, $n=2$ for elliptic flow and $n=3$ for triangular flow).
The goal is to measure the collective anisotropic flow $v_n$ arising from global event-wise correlations to the azimuthal harmonic planes ($\psi_n$), $dN/d\phi\propto 1+\sum_{n=1}2v_n\cos n(\phi-\psi_n)$~\cite{Voloshin:1994mz}.
In the ideal case where collective flow is the only physics, then $V_n=v_n^2$.
However, there exist other physics correlations, such as those from resonance decays, jet-like correlations~\cite{Jacobs:2004qv,Wang:2013qca}, and initial-state gluon correlations~\cite{Dusling:2015gta}, independent of the global event-wise correlations. Those ``genuine" correlations are termed nonflow and contaminate the measured $\mV$ of Eq.~(\ref{eq:V2})~\cite{Borghini:2000cm,Borghini:2006yk,Wang:2008gp,Ollitrault:2009ie}.

Resonance decays and intrajet correlations usually result in particles close in pseudorapidity ($\eta$) and azimuthal angle (dubbed near-side correlations). Their contamination can be reduced by applying an $\eta$-gap between the two POIs or by requiring them to come from phase spaces (subevents) separated in $\eta$~\cite{Poskanzer:1998yz}.
Interjet correlations, or generally momentum conservation effects, usually yield particles back-to-back in azimuth but not locally confined in $\eta$ (dubbed away-side correlations). These nonflow contamination are difficult to remove, and one may resort to model studies or data-driven fitting techniques in 2-dimensional pseudorapidity and azimuthal differences $(\Delta\eta\equiv\eta_1-\eta_2,\Delta\phi\equiv\phi_1-\phi_2)$ of two-particle correlations~\cite{STAR:2023gzg,STAR:2023ioo}.
Because resonance abundances are expected to scale with the final-state multiplicity, nonflow contribution to $\mV$ by Eq.~(\ref{eq:V2}) is approximately proportional to inverse multiplicity. Jet correlations and nuclear effects like jet quenching and baryon-over-meson enhancement may cause nonflow to deviate from such a simple multiplicity scaling.
In general, nonflow contamination is severe in peripheral collisions and become less so towards more central collisions. However, since the collective flow contribution also decreases with increasing centrality because of the more spherical collision zone, nonflow contamination in central collisions can still be appreciable, whereas it is generally the smallest in midcentral collisions.
Experimental data indicate that nonflow contributions can be as large as a few tens of percents in central heavy-ion collisions~\cite{Abdelwahab:2014sge,Feng:2021pgf,STAR:2023gzg,STAR:2023ioo}. 

Despite of a minor contribution to the measured $V_n$ anisotropies, the effect of nonflow can be significant when one is probing small differences in the physics relying on delicate cancellation of final-state effects. For example, relatively large differences in elliptic flow and triangular flow have been observed between central isobar collisions, suggesting significant differences in the deformations of the isobar nuclei~\cite{STAR:2021mii}. It is attempting to extract those deformations directly from their ratio measurements between central isobar collisions, ignoring nonflow contamination, as was done in Ref.~\cite{Zhang:2021kxj}. However, nonflow contamination is important considering the fact that the Ru+Ru/Zr+Zr ratio of $\mV$ is larger than unity whereas the ratio of nonflow is smaller than unity (because of the larger multiplicity in Ru+Ru than in Zr+Zr collisions). The measured isobar ratio of $\sqrt{\mV}$ is 1.025 for full-event and 1.028 for subevent method in the top 0-5\% central isobar collisions~\cite{STAR:2021mii}. Nonflow estimates by $(\Delta\eta,\Delta\phi)$ fit~\cite{STAR:2023gzg,STAR:2023ioo} indicate $\sim$37\% and $\sim$30\% nonflow contamination in the measured $\mV$ by the full-event and subevent methods, respectively (those nonflow contamination are larger than that in central \AuAu\ collisions because the multiplicities are correspondingly smaller). As a result, the isobar ratio of the ``genuine" $v_2$ is 1.046, significantly larger than the measured ratios of $\sqrt{\mV}$~\cite{STAR:2023gzg,STAR:2023ioo}. This would imply that the quadruple deformation difference between Ru and Zr nuclei could be larger than extracted simply from the measured anisotropy difference in Ref.~\cite{Zhang:2021kxj}.

Clearly, nonflow contamination should be removed with well-controlled systematic uncertainties before data can be compared to hydrodynamic model calculations to extract relevant physics quantities. The STAR experiment has recently released a preprint~\cite{STAR:2024eky} extracting the quadruple deformation parameter $\beta_{2\U}$ of the U nucleus by comparing the ratios of the measured $\mv$, $\mdpt$ (where $\delta\pt=\Pt-\mean{\Pt}$), and $\mvpt$, respectively, between two relatively close systems of \UU\ and \AuAu\ collisions to those calculated by hydrodynamic models. The nonflow contamination is assessed by the subevent method.
It is known that the $\eta$-gap or subevent method can only remove part of the nonflow contamination; for example, the away-side nonflow correlations cannot be removed.
It is imperative to examine the effects of remaining nonflow contamination or, more generally, non-hydrodynamic contributions to the measured quantities (e.g., jet contributions to the \vpt\ correlations) on the extracted physics.

To this end, we employ a non-hydrodynamic model, the \hijing\ event generator~\cite{Wang:1991hta,Gyulassy:1994ew} to investigate those effects. \hijing\ is particularly suitable for this study because it is a jet production model with nuclear jet-quenching effects and at the same time contains reasonable descriptions of particle and resonance production. Moreover, it does not have collective flow so the entire azimuthal anisotropy is nonflow; there is no question about the separation of flow and nonflow. 

\section{Simulation and analysis}
The purpose of our study is to examine quantitatively how important nonflow and non-hydrodynamic correlations are to $\mv$, $\mdpt$, and \vpt\ correlations, but not the effects of nuclear deformation of the Uranium (U) or Gold (Au) nuclei. To illustrate the essential point of our study, we keep both the Au and U nuclei spherical. The nuclear densities are given by the Woods-Saxon distributions, $\rho\propto[1+\exp(\frac{r-R}{a})]^{-1}$. The radius parameter $R$ for Au is 6.38~fm and for U is 6.8~fm~\cite{Loizides:2014vua}. The surface diffuseness parameter $a=0.535$~fm is set to be the same for both nuclei. We expect the subtle differences in the density distributions and deformations between the two nuclei will only cause high-order effects, without affecting our main conclusions.

The default \hijing\ (version 1.41) is used. Jet-quenching is switched on. The impact parameter range is set to the maximum of 20~fm to ensure minimum-bias (MB) event samples.
Total $3.8\times10^{8}$ MB events are simulated each for \UU\ collisions at 193~GeV and \AuAu\ collisions at 200~GeV.

The centrality percentiles are determined by the final-state charged hadron multiplicity distributions within pseudorapidity range $|\eta|<0.5$, similar to experiment~\cite{Abelev:2008ab}; the maximum multiplicity corresponds to centrality 0\% and the minimum corresponds to centrality 100\%.

The azimuthal anisotropy is computed by Eq.~(\ref{eq:V2}) averaged over all pairs and all events~\cite{Poskanzer:1998yz}. It can alternatively be computed by averaging over all pairs in an event first and then over all events treating each event equal weighted. The difference is negligible. The $\pt$ fluctuation is also computed by two-particle correlator
\begin{equation}
    \mdpt = \mean{(p_{T,1}-\mean{\pt})(p_{T,2}-\mean{\pt})}\,.
    \label{eq:dpt}
\end{equation}
The alternative calculation by averaging first within event and then over all events, $\mean{[p_{T,1}p_{T,2}]}-\mean{\Pt}^2$, yields negligible difference.
The \vpt\ correlation is obtained by computing event-wise average quantities first and then averaging over all events with equal weight, namely,
\begin{equation}
    \mvpt=\mean{[\cos2(\phi_1-\phi_2)](\Pt-\mean{\Pt})}\,,
\end{equation}
where $[\cos2(\phi_1-\phi_2)]$ stands for the event-wise average of the correlator. Using three-particle cumulant to compute $\mvpt$ yields consistent result because the same particle used in computing $[\cos2(\phi_1-\phi_2)]$ and $\Pt$ does not cause self-correlations.

\section{Simulation results}
\begin{figure}[hbt]
    \includegraphics[width=0.65\linewidth]{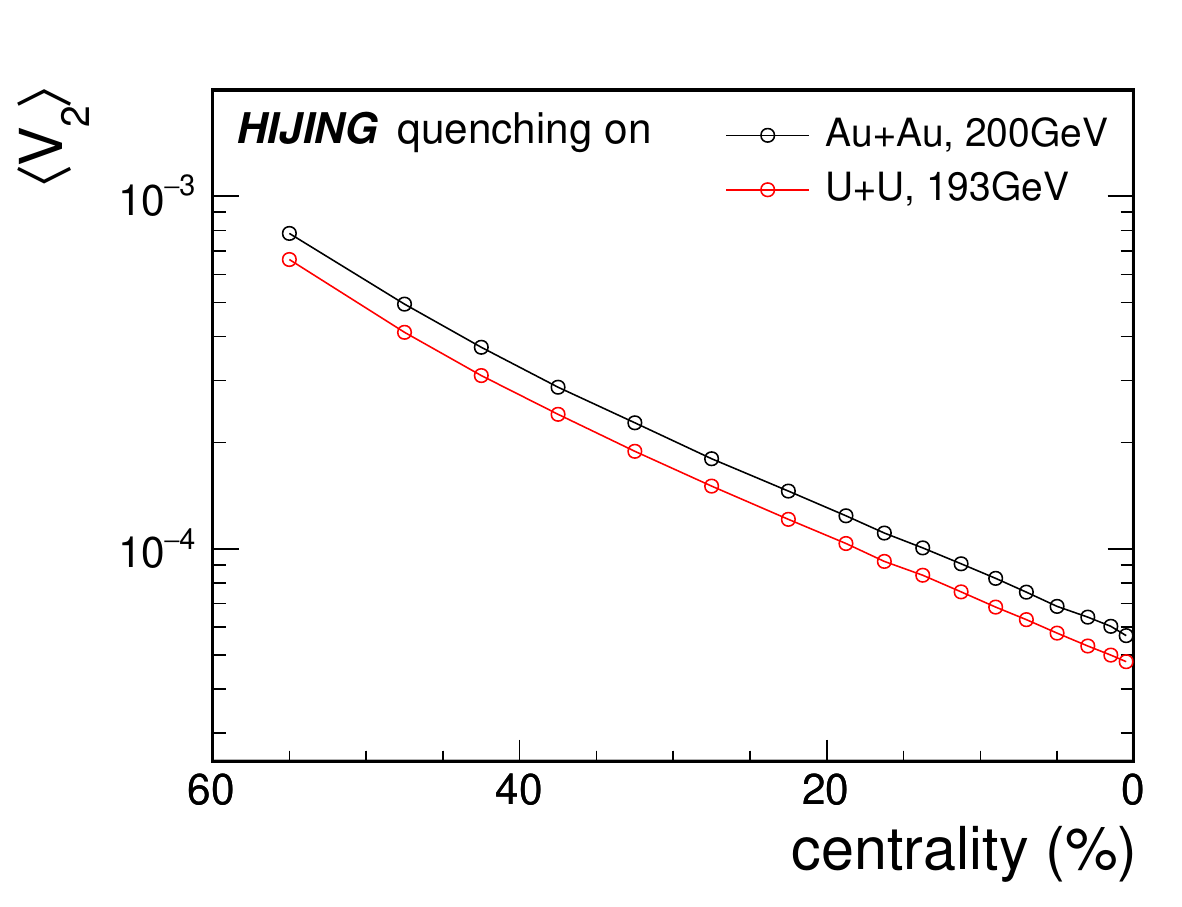}
    \includegraphics[width=0.65\linewidth]{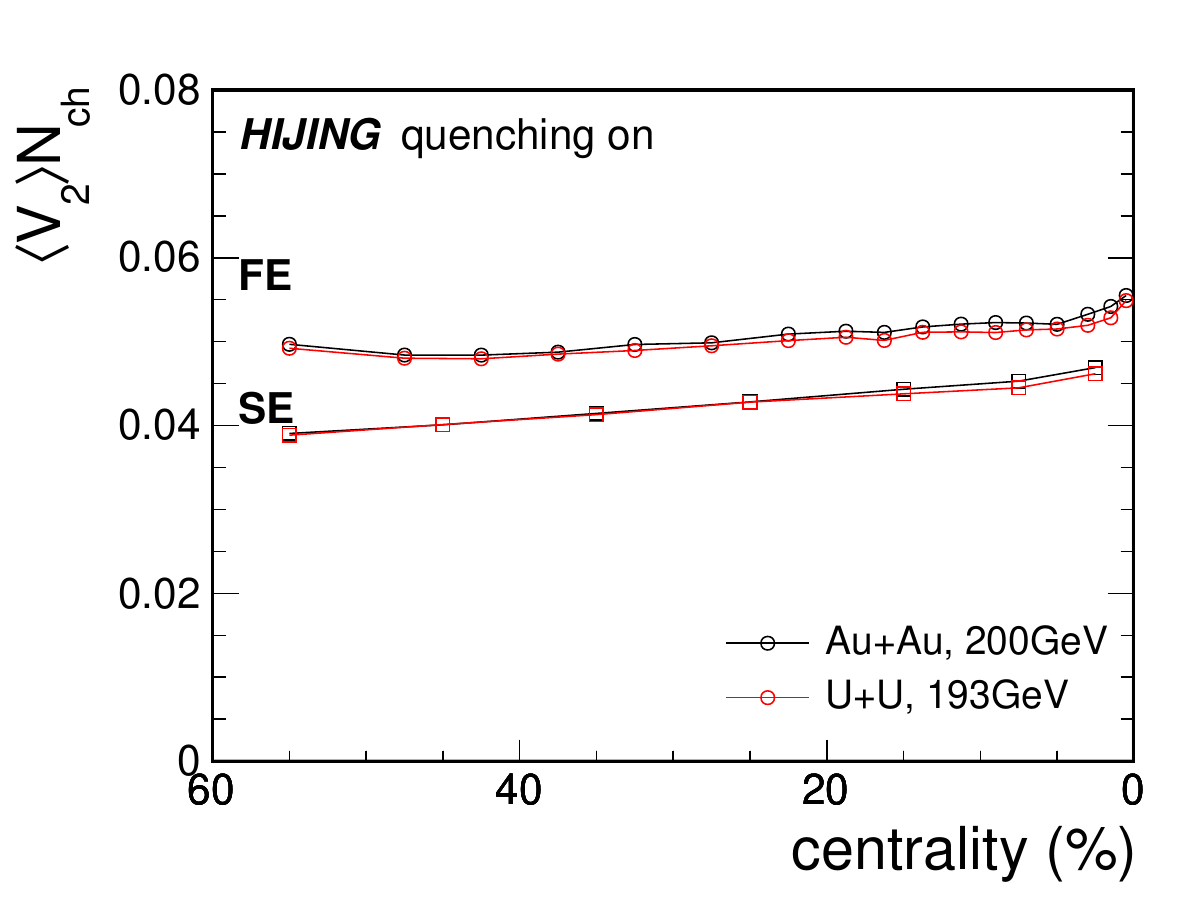}
    \caption{\hijing\ nonflow $\mV$ results. (Upper) Two-particle nonflow $\mV$ of Eq.~(\ref{eq:V2}) as functions of increasing centrality in spherical \AuAu\ (black) and spherical \UU\ (red) collisions. Particles of interest (POI) are taken from pseudorapidity range $|\eta|<1$, and no $\eta$-gap is applied between the two particles (full-event). (Lower) The multiplicity scaled nonflow $\Nch\mV$ from the full-event (FE, circles) method, as well as from the subevent (SE, squares) method where one POI is taken from $-1<\eta<-0.1$ and the other from $0.1<\eta<1$.} 
    \label{fig:V2}
\end{figure}

Figure~\ref{fig:V2} upper panel shows the nonflow correlations in \hijing\ calculated by Eq.~(\ref{eq:V2}) as functions of centrality from 60\% to 0\% in \AuAu\ and \UU\ collisions. The POIs are both taken from $|\eta|<1$ (full-event method). The nonflow in \UU\ is lower than that in \AuAu\ by approximately 20\%, the multiplicity difference at a given centrality. 
In the top 0-5\% centrality, the \hijing\ nonflow correlations are $\mV\approx 0.6\times10^{-4}$ in \AuAu\ and $0.5\times10^{-4}$ in \UU\ collisions, respectively.

The lower panel of Fig.~\ref{fig:V2} shows the charged hadron multiplicity ($\Nch$) scaled nonflow correlations, $\Nch \mV$, for both full-event and subevent methods. The subevent method takes one POI from $-1<\eta<-0.1$ and the other from $0.1<\eta<1$. 
The $\Nch\mV$ values are consistent between the two systems, indicating the same nonflow sources diluted by multiplicity.
The $\Nch\mV$ values show only a minor increase with centrality, in line with the approximate inverse multiplicity scaling of nonflow. The minor increase may be attributed to final-state nuclear effects like jet-quenching and enhanced heavy resonance production. The reduction in nonflow from the full-event to subevent method is small, only 12\% in the most central collisions. This implies that the away-side correlations make a large contribution to the overall nonflow, the near-side correlations are broad and the $\eta$-gap applied by the subevent method removes only part of the near-side nonflow, or likely both.

\begin{figure}[hbt]
    \includegraphics[width=0.65\linewidth]{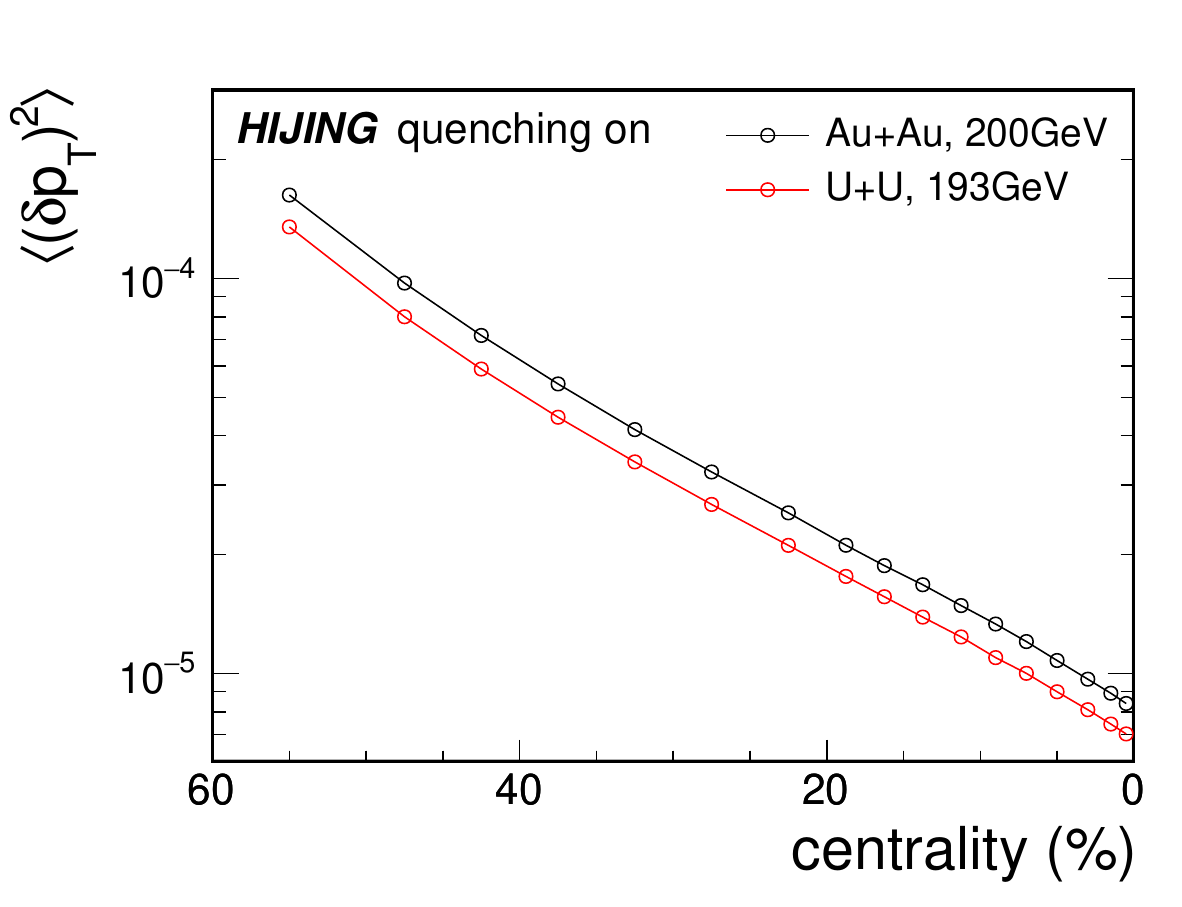}
    \includegraphics[width=0.65\linewidth]{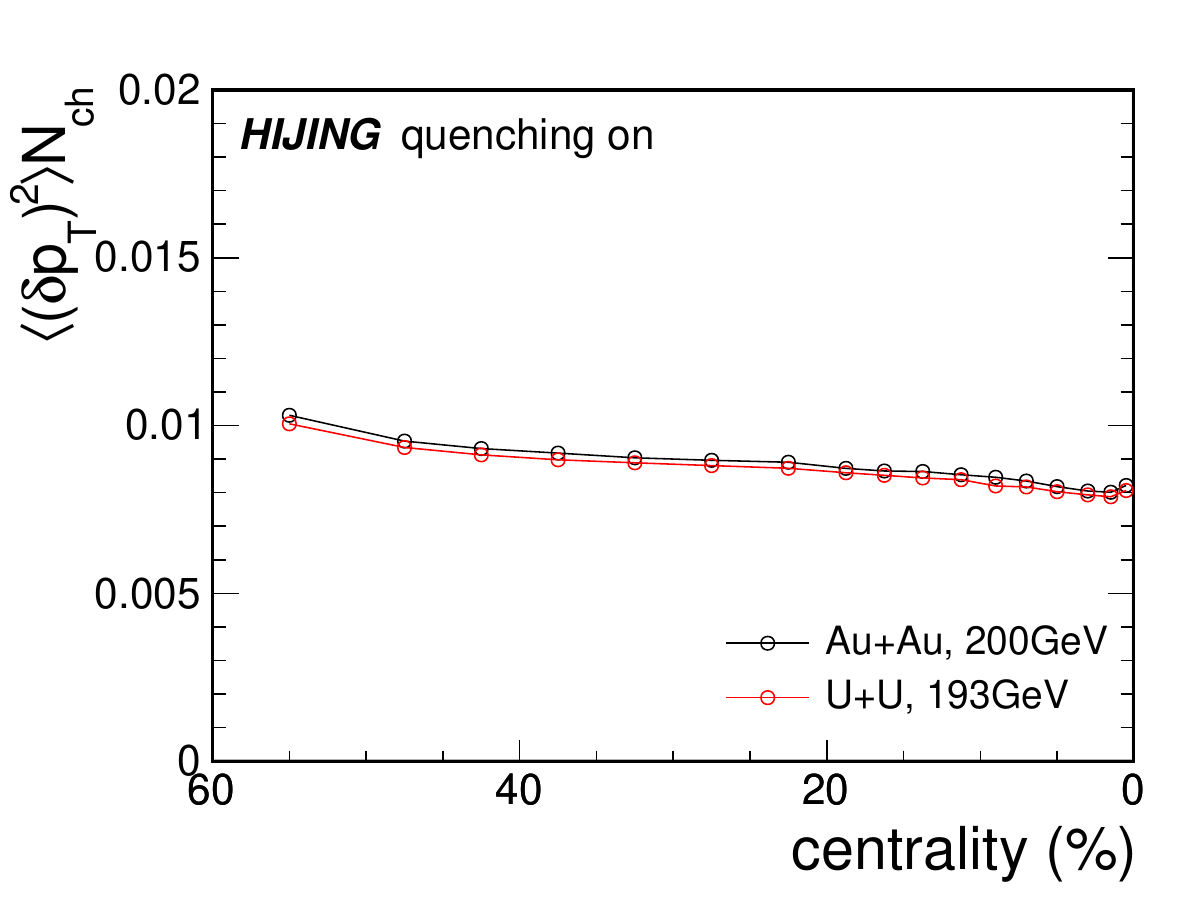}
    \caption{\hijing\ $\pt$ correlation $\mdpt$ results. Event-wise $\pt$ fluctuations as functions of centrality in spherical \AuAu\ (black) and spherical \UU\ (red) collisions. The POIs are taken from pseudorapidity range  $|\eta|<1$, and no $\eta$-gap is applied between the two particles (full-event).}
    \label{fig:dpt}
\end{figure}

Figure~\ref{fig:dpt} shows the $\pt$ fluctuations calculated by Eq.~(\ref{eq:dpt}) with the full-event method in \AuAu\ and \UU\ collisions as functions of centrality. The $\mdpt$ values are approximately 20\% smaller in \UU\ than in \AuAu\ collisions, similar to the $\mV$ results in Fig.~\ref{fig:V2}. This is because both quantities are essentially two-particle correlation measures and are diluted similarly by multiplicity. The lower panel of Fig.~\ref{fig:dpt} shows the multiplicity scaled $\pt$ fluctuations, $\Nch\mdpt$. Indeed, the $\Nch\mdpt$ values equal between \AuAu\ and \UU\ collisions at a given centrality. The $\Nch\mdpt$ value exhibits a  modest decreases  with increasing centrality.

\begin{figure*}
    \includegraphics[width=0.325\linewidth]{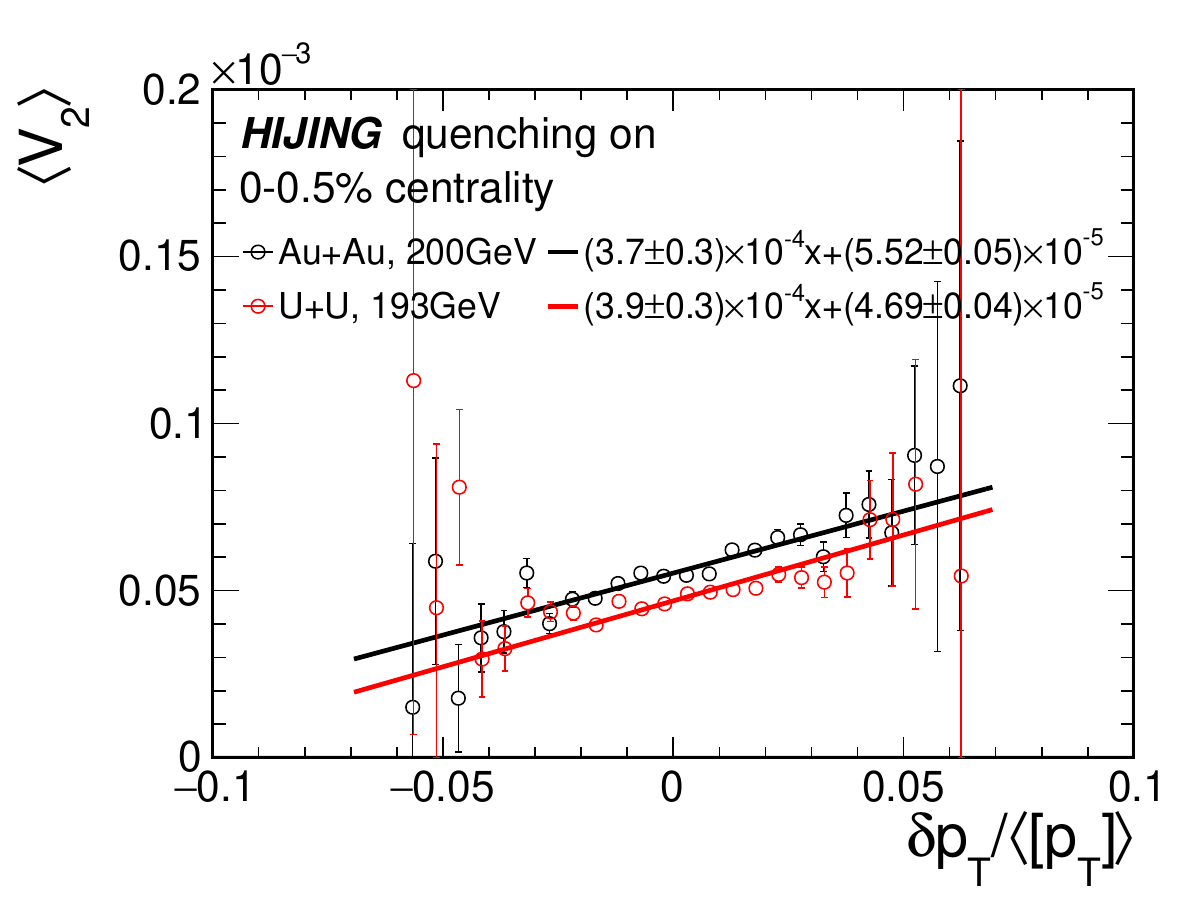}
    \includegraphics[width=0.325\linewidth]{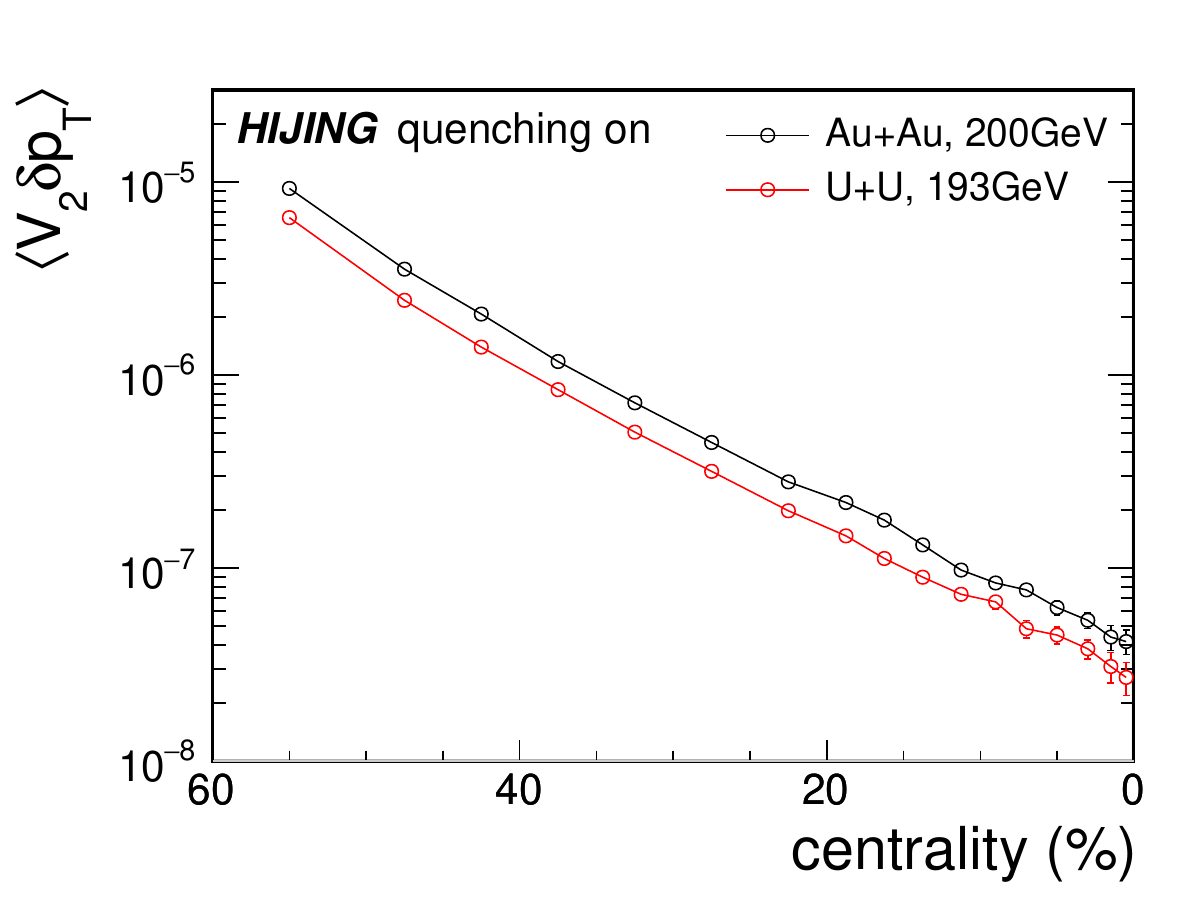}
    \includegraphics[width=0.325\linewidth]{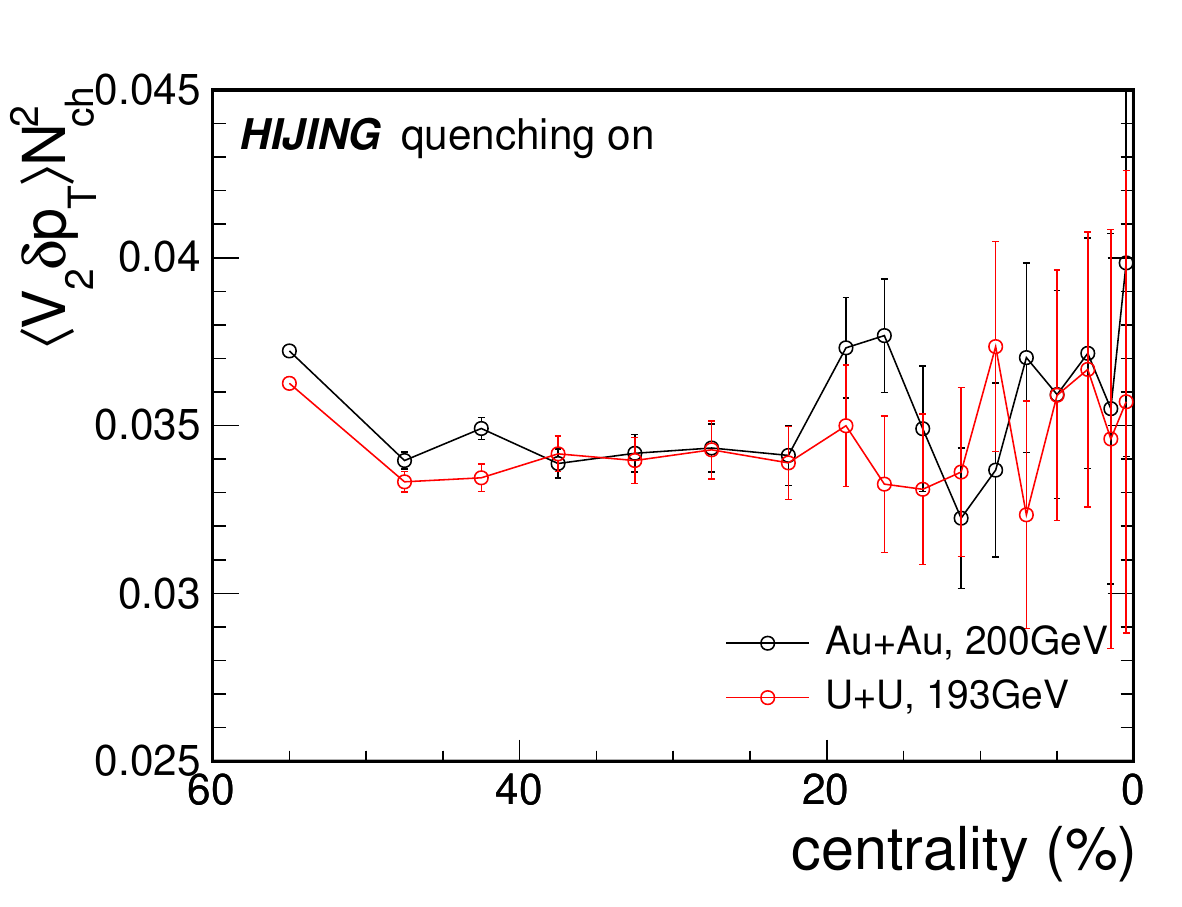}
    \caption{\hijing\ \vpt\ correlation results.
     Event-by-event correlations between $\mV$ and $\delta\pt/\mean{\Pt}$ in the top 0-0.5\% centrality (left panel), correlation covariance $\mvpt$ as functions of centrality (center panel), and the multiplicity scaled $\Nch^2\mvpt$ (right panel) in spherical \AuAu\ (black) and spherical \UU\ (red) collisions. The POIs are taken from pseudorapidity range  $|\eta|<1$, and no $\eta$-gap is applied between the two particles (full-event).}
    \label{fig:v2pt}
\end{figure*}

Figure~\ref{fig:v2pt} left panel shows the event-by-event correlations between $\mV$ and $\delta\pt/\mean{\Pt}$ calculated by \hijing\ in the top 0-0.5\% centrality for \AuAu\ and \UU\ collisions. The correlations are positive and the correlation strengths appear to be similar in terms of the slope of $\mV$ vs.~$\delta\pt/\mean{\Pt}$. This can be understood by minijets which contribute to both nonflow and $\pt$ fluctuations and the physics is no different between \AuAu\ and \UU. 
The overall value of $\mV$ is smaller in \UU\ than in \AuAu\ because of the multiplicity difference (see Fig.~\ref{fig:V2}), so the $\mvpt$ value is smaller.

The center panel of Fig.~\ref{fig:v2pt} shows the \vpt\ correlation quantity $\mvpt$ in \AuAu\ and \UU\ collisions as functions of centrality. Similar to $\mV$ and $\mdpt$, the \UU\ system shows a smaller correlation magnitude than the \AuAu\ system, by approximately 30-40\%. This is in line with the smaller magnitudes of $\mV$ and $\mdpt$ in \UU\ than in \AuAu\ collisions, each by approximately 20\% as shown in Figs.~\ref{fig:V2} and~\ref{fig:dpt}, respectively. Indeed, the $\Nch^2$ scaled \vpt\ correlations are similar between the two systems as shown in the right panel of Fig.~\ref{fig:v2pt}.

It is clear that $\mV$, $\mdpt$, and $\mvpt$ correlations are all nonzero in \hijing. These correlations are all caused by non-hydrodynamic origins because \hijing\ is not a hydrodynamic model. \hijing\ describes jet production in hadron-hadron, hadron-nucleus, and nucleus-nucleus collisions, and jet quenching by partonic energy loss in the final-state nuclear medium~\cite{Wang:1991hta,Gyulassy:1994ew}. \hijing\ has also reasonable descriptions of soft particle and resonance production via string fragmentation. Therefore, the $\mV$, $\mdpt$, and $\mvpt$ correlations observed in \hijing\  presumably arise mainly from jet correlations and resonance decays.

\section{Implications to STAR data and Discussions}
In the recent preprint~\cite{STAR:2024eky}, the STAR experiment compared ratios of correlation quantities in \UU\ collisions at 193~GeV over \AuAu\ collisions at 200~GeV  to those calculated by the IP-Glasma+MUSIC hydrodynamic model~\cite{Schenke:2020uqq,Schenke:2020mbo} to extract the deformation parameters of the U nucleus. From the ratio of the measured $\mv$, the extracted quadruple deformation parameter is $\beta_{2\U}=0.240\pm0.014$; this value is, however, discounted attributing to an inadequate description of $v_2$ by hydrodynamics (cf Figure~3a of Ref.~\cite{STAR:2024eky}). From the simultaneous comparisons of the measured $\mdpt$ ratio and $\mvpt$ ratio to hydrodynamic calculations, the quadruple deformation and triaxiality parameters are extracted to be $\beta_{2\U}=0.297\pm0.013$ and $\gamma_{\U}=8.6^\circ\pm4.7^\circ$, respectively. Comparing to another hydrodynamic model (trajectum~\cite{Nijs:2023yab}) leads to a value of $\beta_{2\U}=0.273\pm0.015$. 
This value is not covered by the quoted systematic uncertainty of the main $\beta_{2\U}$ result.

The nonflow contamination in $\mv$ is assessed in the STAR work~\cite{STAR:2024eky} as follows: the $\mv$ is taken to be the average of the full-event measurement where POIs are taken from the STAR Time Projection Chamber (TPC) acceptance ($|\eta|<1$) and the subevent measurement where the POIs are from separate subevents ($-1<\eta<-0.1$ and $0.1<\eta<1$), and half of their difference is taken to be the systematic uncertainty. 
The $\mdpt$ and $\mvpt$ are measured similarly by full-event and subevent methods, and their systematic uncertainties are assessed in similar ways. 
The assessed {\em nonflow} systematic uncertainties are 1-2\% for $\mv$, 1-3\% for $\mdpt$, and 2-4\% for $\mvpt$, with the corresponding {\em total} systematic uncertainties to be 2.5-4\%, 2-5\%, and 4-10\%, respectively. 

Figure~\ref{fig:V2} shows that $\mV=0.6\times10^{-4}$ in the most central 0-5\% \AuAu\ collisions in \hijing. This is about 12\% of the measured $\mv=5\times10^{-4}$~\cite{STAR:2024eky} in the same centrality range of \AuAu\ collisions.
According to \hijing, only 12\% of it (i.e., 12\% of the 12\% nonflow) is removed by the subevent method, as shown in the lower panel of Fig.~\ref{fig:V2}. This implies that significant nonflow remains in the STAR $\mv$ measurement, and the nonflow uncertainty is severely underestimated in~\cite{STAR:2024eky}, possibly by an order of magnitude (a factor of 15 based on \hijing\ face value).
Because nonflow approximately scales with the inverse multiplicity which differs by 20\% between \AuAu\ and \UU\ collisions, a 12\% remaining nonflow would yield a 2\% effect in the ratio of $R_{v_2^2}=\mv_{\rm U}/\mv_{\rm Au}$. 
The real situation is worse because the $v_2$ in central \UU\ collisions is larger due to its deformity (see Figure~4a of Ref.~\cite{STAR:2024eky}), so the nonflow contamination in central \UU\ is significantly smaller than in \AuAu, making the effect on the ratio $R_{v_2^2}$ much larger (this point is more quantitatively presented later). 

It is conceivable that non-hydrodynamic correlations in $\pt$-$\pt$ and \vpt\ correlations cannot be completely removed by the subevent method either. Part of those non-hydrodynamic correlations are due to minijets which are wide objects comparable to the STAR TPC acceptance; furthermore, dijet correlations go far beyond the acceptance. It is reasonable to conjecture that majority of those non-hydrodynamic correlations remain in $\mdpt$ and $\mvpt$, like in $\mv$. This implies that the uncertainties due to nonflow and non-hydrodynamic correlations could be much larger than estimated in~\cite{STAR:2024eky}, easily reaching 10\% or beyond. 

\begin{figure*}[hbt]
    \begin{minipage}{0.305\textwidth}
    \includegraphics[width=\textwidth]{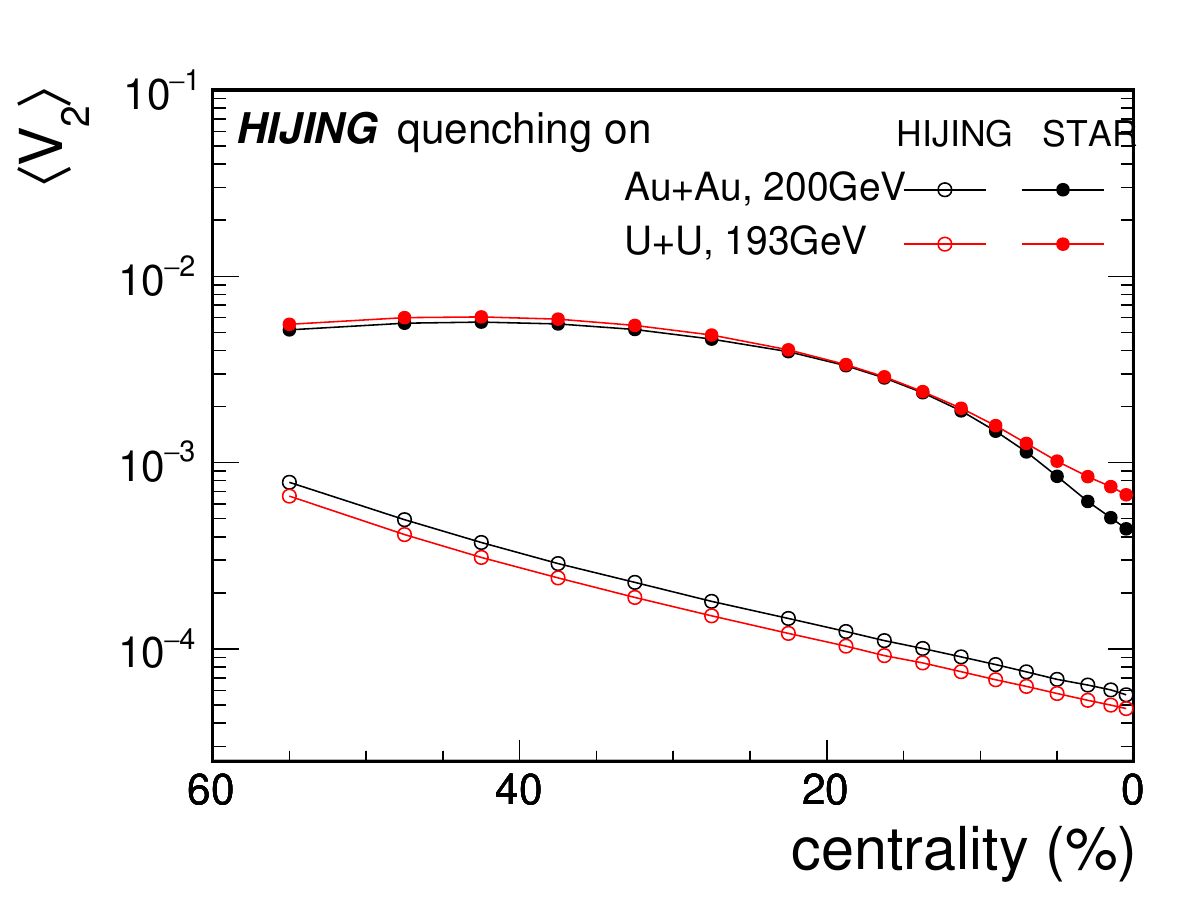}\vspace{-0.1cm}
    \includegraphics[width=\textwidth]{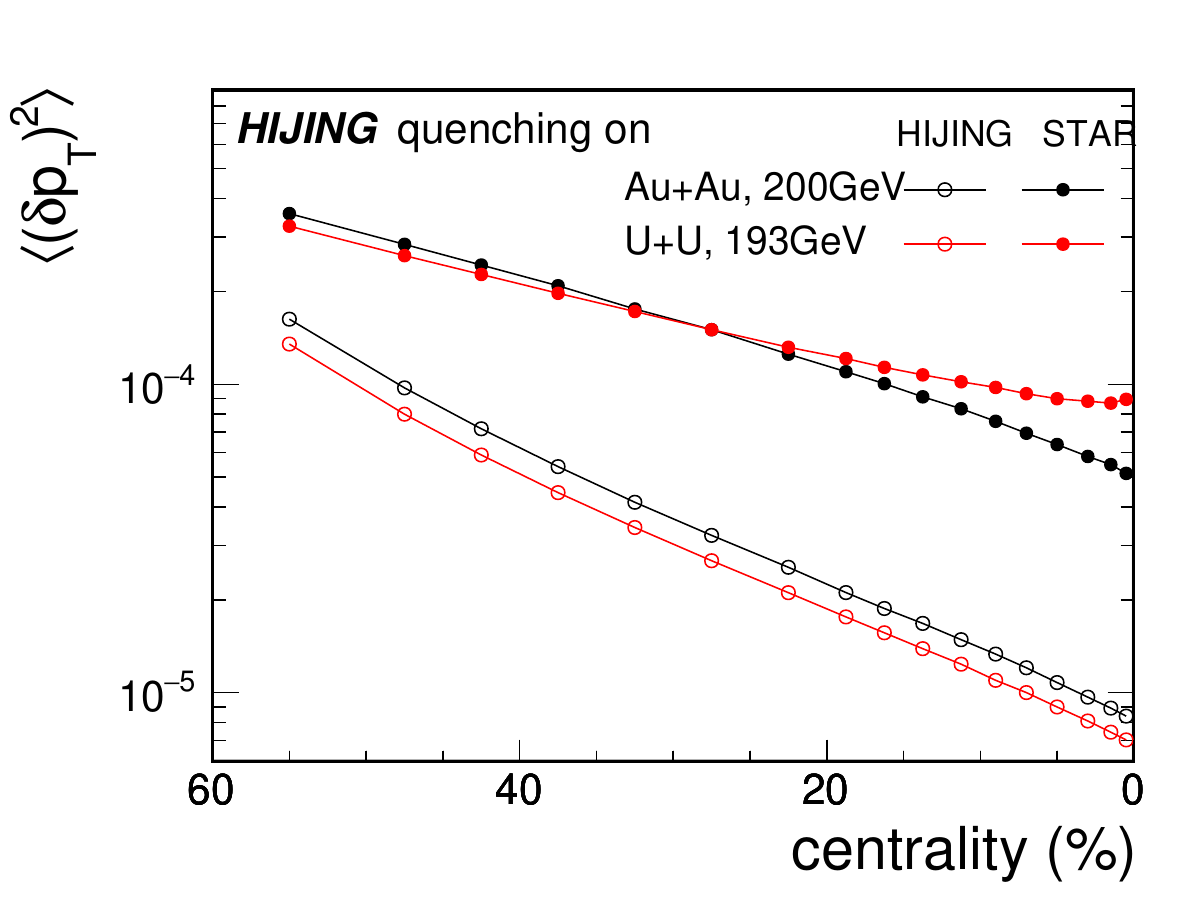}\vspace{-0.1cm}
    \includegraphics[width=\textwidth]{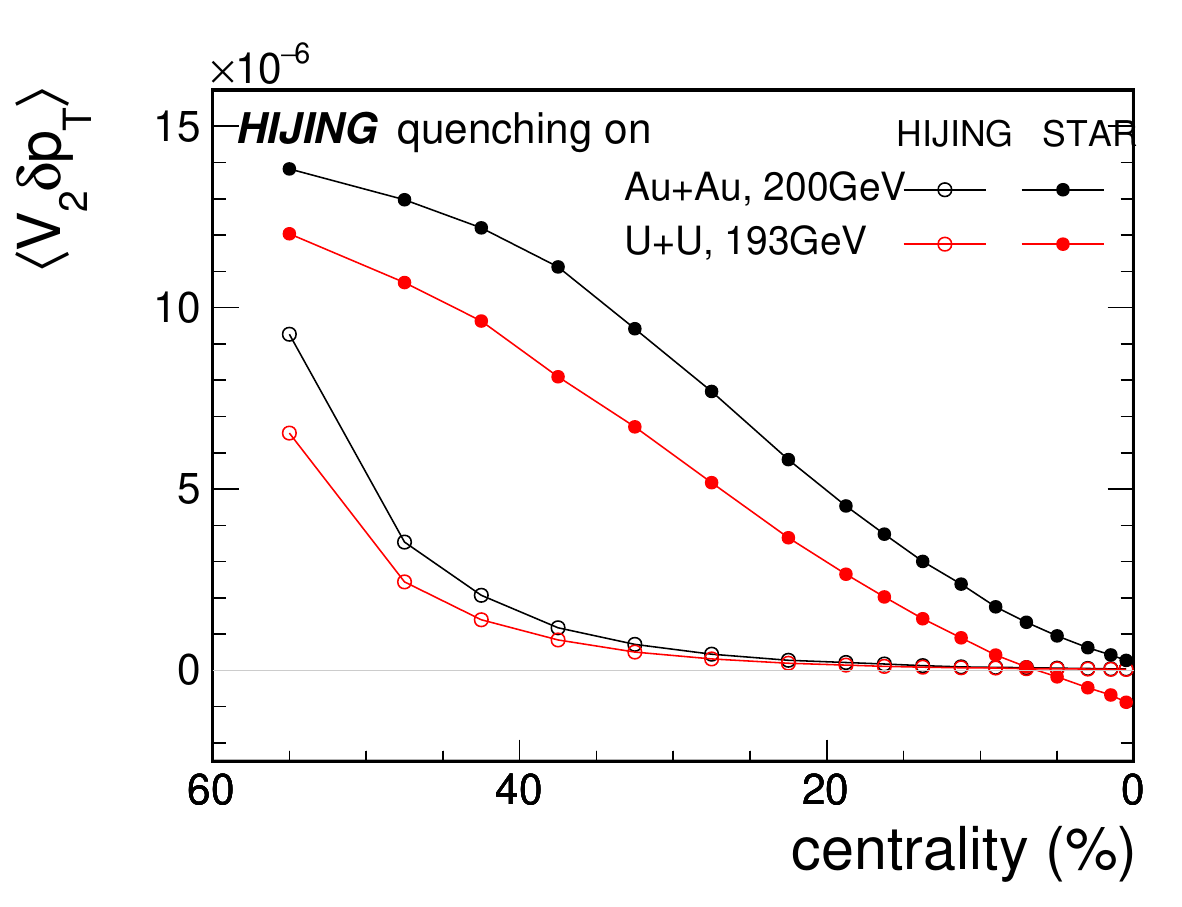}
    \end{minipage}
    \begin{minipage}{0.685\textwidth}
    \vspace*{0.4cm}
    \includegraphics[width=\textwidth]{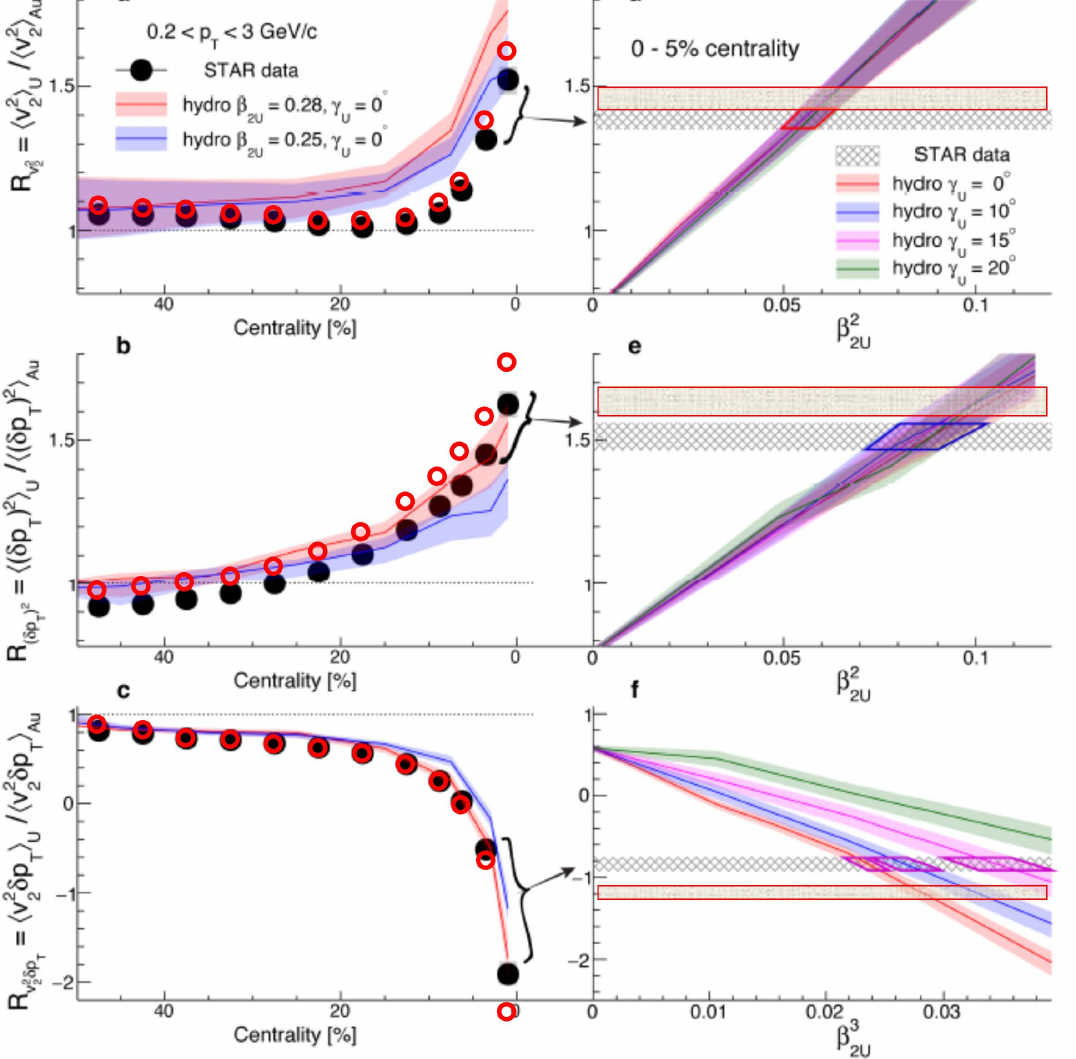}
    \end{minipage}   
    \caption{Effects of non-hydrodynamic correlations on STAR data. (Left panels) The $\mv$, $\mdpt$, and $\mvpt$ correlations calculated by \hijing\ (open points) compared to STAR data~\cite{STAR:2024eky} (filled points). The \hijing\ results are from full-event method and the STAR data are from the subevent method. The \hijing\ subevent result are relatively 12\% smaller than the full-event result for $\mv$, and likely similar for $\mdpt$ and $\mvpt$. (Middle and right panels) Copy of part of Figure~3 from Ref.~\cite{STAR:2024eky} with hand-drawn illustrations of how the STAR data points would move and the extracted U nucleus deformation parameters would change if non-hydrodynamic effects were corrected according to \hijing.}
    \label{fig:comp}
\end{figure*}

Since nonflow can be corrected in many analyses, including this recent STAR work~\cite{STAR:2024eky}, it should rather be treated as a systematic error and be corrected, not simply a systematic uncertainty. Earnest efforts should be spent to correct for the systematic errors and assess the systematic uncertainties on such corrections. With this in mind, we compare in Fig.~\ref{fig:comp} left panels our \hijing\ calculations of the quantities $\mv$, $\mdpt$, and $\mvpt$ to those measured by STAR~\cite{STAR:2024eky}. In the middle and right panels of Fig.~\ref{fig:comp}, we illustrate how the measured \UU/\AuAu\ ratios of these quantities would move and how the extracted deformation parameters would change if nonflow and non-hydrodynamic effects were corrected according to \hijing. 

Figure~\ref{fig:comp} left-top panel shows that the measured $\mv$ values in the top 0-5\% centrality are approximately $5\times10^{-4}$ for \AuAu\ and $8\times10^{-4}$ for \UU. The significantly larger $\mv$ measured for \UU\ is due to the large quadruple deformation. Thus, the \hijing\ nonflow contributions to $\mv$ 
would be 12\% in central \AuAu\ and 6\% in central \UU\ collisions. This would result in an increased $R_{v_2^2}$ by 6\% in the top 0-5\% centrality, in contrast to the 2\% from simple multiplicity scaling aforementioned. This is illustrated in the center- and right-top panels of Fig.~\ref{fig:comp}. Subtracting \hijing\ nonflow from data would raise the STAR data points to the red circles in the center-top panel, and the hatched area of the top 0-5\% data to the red box in the right-top panel. This is about 6 times the estimated systematic uncertainty on $\mv$ due to nonflow in~\cite{STAR:2024eky}. It is not quite as large as the factor of 15 mentioned earlier, possibly implying that nonflow in \hijing\ is an underestimate for real data and/or the subevent method is less effective in \hijing. 
Indeed, the estimated nonflow in the full-event $v_2^2$ data is on the order of 20\% in central \AuAu\ collisions~\cite{Abdelwahab:2014sge}.

Figure~\ref{fig:comp} left-middle panel shows that $\mdpt$ from \hijing\ is also appreciable compared to the STAR data. The \UU\ data increase towards central collisions presumably because of its deformity. 
In the most central 0-5\% collisions, the $\mdpt$ value in \hijing\ is $\sim$20\% of that measured in \AuAu, while that for \UU\ is $\sim$10\%. 
Correcting for these non-hydrodynamic effects would increase the ratio of $R_{(\delta\pt)^2}=\mdpt_{\rm U}/\mdpt_{\rm Au}$ by approximately 10\% in the top 0-5\% centrality, again several times the estimated systematic uncertainty due to nonflow in~\cite{STAR:2024eky}. This is illustrated in the center- and right-middle panels of Fig.~\ref{fig:comp}, similarly to the corresponding top panels.

Figure~\ref{fig:comp} left-bottom panel shows that the $\mvpt$ correlations in \hijing\ are strong in the 60-50\% centrality range and rapidly decrease towards central collisions. The decrease in the measured data is not as rapid. For the top 0-5\% central \AuAu\ collisions, $\mvpt$ from \hijing\ is approximately $5\times10^{-8}$, 
about 10\% of that measured in data. In central \UU\ collisions, the measured $\mvpt$ correlations become significantly negative, presumably because of  hydrodynamic responses to the large deformation of the U nucleus. The $\mvpt$ correlations from \hijing\ remain positive in the top 0-5\% central \UU\ collisions, approximately $3.5\times10^{-8}$. 
This is about 6\% of the magnitude of the measured negative \vpt\ correlations in 0-5\% central \UU\ collisions. The positive correlations in \hijing\ likely arise from fluctuations in resonance $\pt$ and jet production. For the former, the larger the resonance $\pt$, the smaller the decay opening angle, and thus the larger the nonflow $V_2$. For the latter, the more the jet production, the larger the $\Pt$ and the larger the nonflow $V_2$. The overall effect of $\mvpt$ correlations in \hijing\ would cause a more negative \UU/\AuAu\ ratio of $\mvpt$ by approximately 16\% in the top 0-5\% centrality, again several times the estimated systematic uncertainty due to nonflow in~\cite{STAR:2024eky}. This is illustrated in the center- and right-bottom panels of Fig.~\ref{fig:comp}. 

Note that in our simulations spherical nuclei are used for both Au and U. Deformed nuclei may enhance fluctuations in $\pt$ and in jet production, the effects of which are likely minor. On the other hand, nuclear deformation would not generate a negative \vpt\ correlations in \hijing\ as the physics underlying those correlations are jets and resonance decays; hydrodynamic physics causing the \vpt\ anti-correlations is not present in \hijing. A larger \vpt\ correlations in \UU\ due to deformation would cause the nonflow correction to be even more negative for $\mvpt$, whereas a larger $\pt$ fluctuations in \UU\ would reduce the correction for $\mdpt$. 

It is worth to emphasize that the nonflow and non-hydrodynamic correlations in \hijing\ are several times the nonflow systematic uncertainties estimated in Ref.~\cite{STAR:2024eky}. Correcting for those non-hydrodynamic correlations calculated by \hijing\ would result in a quadruple deformation parameter $\beta_{2\U}$ of the U nucleus different from that obtained by STAR, even beyond  the quoted {\em total} uncertainty~\cite{STAR:2024eky} (see the right panels of Fig.~\ref{fig:comp}). 
Those total systematic uncertainties include not only those on nonflow but also all other sources such as experimental systematics from analysis cut variations and  theoretical systematics from hydrodynamic model parameter dependencies~\cite{STAR:2024eky}. 
We thus conclude that the extracted deformation parameter $\beta_{2\U}$ from relativistic heavy-ion collisions in Ref.~\cite{STAR:2024eky} is premature.

It is noteworthy that \hijing\ is only one particular model in describing relativistic heavy-ion collisions, primarily for jet production and jet quenching. It is used here mainly to illustrate the potential significance of nonflow contamination; it is not meant to be {\em the} correction for experimental data, so no attempt is made to assess systematic uncertainties on such a correction by \hijing. The red boxes drawn in the right panels of Fig.~\ref{fig:comp} are merely the shaded area of the STAR data shifted by the \hijing\ corrections, keeping the total  uncertainties the same. 

In real data analysis, a more thorough and rigorous study would be required to correct for nonflow contamination and assess the systematic uncertainties on such corrections, for example, by investigating not only \hijing\ but also other heavy-ion models and by rigorous estimation of nonflow in real data. One data-driven estimate was performed by STAR exploiting the $\eta$ reflection symmetry in symmetric heavy-ion collisions; the estimated nonflow in central \AuAu\ collisions is on the order of 20\%~\cite{Abdelwahab:2014sge}. Another viable data-driven technique is to perform 2-dimensional $(\Delta\eta,\Delta\phi)$ fits to two-particle correlations~\cite{STAR:2006lbt,STAR:2011ryj,STAR:2023gzg,STAR:2023ioo}; a recent such study indicates an approximately 40\% nonflow in central isobar collisions~\cite{STAR:2023gzg,STAR:2023ioo}, in line with the central \AuAu\ data considering multiplicity dilution of nonflow. Assuming inverse multiplicity scaling of nonflow over the entire centrality from peripheral to central collisions, the 70-80\% peripheral \AuAu\ measurement ($v_2^{\rm peri}\approx6.9\%$), if taken as all nonflow, would constitute a 30\% nonflow, i.e., $(v_2^{\rm peri})^2\frac{\Nch^{\rm peri}}{\Nch^{\rm cent}}$, in the 0-5\% central collision measurement ($v_2^{\rm cent}\approx2.4\%$)~\cite{Adams:2004bi}. 
Comparing the accumulative correlations as functions of $\pt$ in the top 0-5\% central \AuAu\ to MB proton-proton collisions suggests a nonflow contribution on the order of 15\% in the former~\cite{STAR:2004amg}. 
All these estimates indicate an appreciable nonflow contribution in central \AuAu\ collisions.

\section{Conclusions}
It is well known that anisotropic flow measurements are contaminated by nonflow correlations from resonance decays, jet correlations, etc. In general, non-hydrodynamic correlations can give rise to many types of correlations including $\pt$ fluctuations and \vpt\ correlations. There is absolutely no question about those correlations being present and relatively large, as seen in \hijing, a non-hydrodynamic model with no flow.

In the recent preprint by STAR~\cite{STAR:2024eky}, elliptic flow $v_2$, $\pt$ fluctuations, and \vpt\ correlations were measured in \UU\ and \AuAu\ collisions and their ratios in the most central collisions were directly compared to hydrodynamic calculations, with minimal estimation of nonflow contamination, to extract an Uranium quadruple deformation parameter $\beta_{2\U}=0.297\pm0.013$ of high significance. We examine non-hydrodynamic contributions to those correlation quantities using the \hijing\ model and demonstrate that those contributions could present significant contamination to the STAR measurements. Correcting for those contamination according to \hijing\ would yield a $\beta_{2\U}$ value beyond the  quoted  total uncertainty. 
We conclude that the nonflow contamination is inadequately estimated in the STAR measurements~\cite{STAR:2024eky} and the extracted $\beta_{2\U}$ parameter is deemed premature. Additional homework is needed to put it onto a firm ground.

Our study is motivated by the recent STAR work at RHIC, but our message goes beyond. The most important message of our study is that nonflow and non-hydrodynamic correlations should be earnestly assessed with faithful systematic uncertainties assigned, either after correction wherever possible or without correction, before drawing physics conclusions from comparing experimental data to hydrodynamic calculations.

\section*{Acknowledgment} 
This work is supported in part by the U.S.~Department of Energy (Grant No.~DE-SC0012910).

\bibliographystyle{unsrt}
\bibliography{ref}

\end{document}